\documentclass[11pt,a4paper]{article}
\usepackage{jheppub_kim}

\usepackage{subfigure,amsmath,amssymb ,amsfonts, latexsym}

\usepackage[notcite,color,final]{showkeys}
\definecolor{refkey}{gray}{.25}
\definecolor{labelkey}{gray}{.25}

\newcommand{\be}{\begin{eqnarray}}
\newcommand{\ee}{\end{eqnarray}}

\def\ben{\begin{equation}}
\def\een{\end{equation}}
\def\bena{\begin{eqnarray}}
\def\eena{\end{eqnarray}}

\newcommand{\bn}{\bar{\nabla}}

\newcommand{\cD}{{\cal{D}}}
\newcommand{\hD}{{\hat{D}}}

 \newcommand{\dotal}{\dot{\alpha}}
 
 \newcommand{\dotga}{\dot{\gamma}}

\newcommand{\beq}{\begin{equation}}
\newcommand{\eeq}{\end{equation}}
\newcommand{\bea}{\begin{eqnarray}}
\newcommand{\eea}{\end{eqnarray}}

\newcommand{\N}{{\cal{N}}}
\newcommand{\F}{{\cal{F}}}
 
\newcommand{\bean}{\begin{eqnarray*}}
\newcommand{\eean}{\end{eqnarray*}}
\newcommand{\bs}{\begin{subequations}}
\newcommand{\es}{\end{subequations}}

   \vspace*{0cm}

\subheader{LMU-ASC 09/12}
\title{A New Class of Four-Dimensional $\N=1$ Supergravity with Non-minimal Derivative Couplings}

\author[a]{Fotis Farakos}
\author[b]{Cristiano Germani}
\author[a]{Alex Kehagias}
\author[a,c]{Emmanuel N. Saridakis}

\affiliation[a]{Physics Division, National Technical University of Athens,
15780 Zografou Campus,  Athens, Greece}

\affiliation[b]{Arnold Sommerfeld Center, Ludwig-Maximilians-University,
Theresienstr. 37, 80333 Muenchen, Germany}

\affiliation[c]{CASPER, Physics Department, Baylor University, Waco, TX 
76798-7310, USA}

\emailAdd{fotisf@mail.ntua.gr}
\emailAdd{ cristiano.germani@lmu.de}
\emailAdd{kehagias@central.ntua.gr}
\emailAdd{Emmanuel$_-$Saridakis@baylor.edu}

\keywords{nonminimal derivative coupling,
supergravity}

\abstract{In the  $\N=1$ four-dimensional new-minimal
supergravity framework, we 
supersymmetrise the coupling of the scalar kinetic term 
to the Einstein tensor. This coupling, although 
introduces a non-minimal derivative 
interaction of
curvature to matter, it does not introduce harmful higher-derivatives.
For this construction, we employ off-shell chiral and real linear
multiplets. Physical scalars are accommodated
 in the 
 chiral multiplet whereas curvature resides in a linear
 one.}

\begin{document}

\maketitle

\section{Introduction}

The most generic theory propagating a massless spin-2 and a scalar degree of freedom is {\it not} 
General Relativity minimally coupled to a scalar field (GRM). 
Indeed,  Horndeski \cite{hord} proved that tensor-scalar theories with only second order differential
 equations are not restricted to GRM. Up to quadratic terms in matter fields and in four-dimensions, Horndeski showed that the most generic theories propagating a massless spin-2 and a spin-0 are
\be\label{theory}
{\cal L}={\cal L}_{\rm GRM} \pm \frac{1}{M_I^2}{\cal{L}}_{I\phantom{I}}\pm\frac{1}{M_{II}^2}{\cal{L}}_{II}+\xi{\cal{L}}_{III}\ ,
\ee
where
\be
{\cal L}_{\rm GRM}&=&\frac{1}{2}\left[M_P^2 R-\partial_a
\phi\partial^a\phi\right],\\
 {\cal{L}}_{I\phantom{I}}&=&
\left(M_{\phi}^I\phi+\phi^2\right) R_{GB}^2\, ,\label{GB}\\
{\cal{L}}_{II}&=& G^{\mu\nu}\partial_\mu \phi\partial_\nu  \phi\, ,
\label{Gf}\\
{\cal{L}}_{III}&=&\left(M_\phi^{III}\phi+ \phi^2\right)R\, ,
\ee
and
\be
G_{\mu\nu}=R_{\mu\nu}-\frac{1}{2} g_{\mu\nu}R\, , ~~~ R_{GB}^2=R_{\mu\nu\gamma\delta}R^{\mu\nu\gamma\delta}
-4 R_{\mu\nu} R^{\mu\nu}+R^2
\ee  
are  the Einstein and Gauss-Bonnet tensors, respectively, $M_{(I,II)}, M_\phi^{I,II}$ are mass scales, $\xi$ a constant and finally $M_P$ is the Planck constant. That ${\cal{L}}_{I\phantom{I}}$ leads to second order evolution
equation follows easily from the fact that the Gauss-Bonnet combination is a total derivative in four-dimensions and it is linear in second order derivatives. Instead, ${\cal{L}}_{II}$ leads to second order equations as, in Hamiltonian ADM formalism \cite{adm}, $G_{tt}$ and $G_{it}$ contain only first time
derivatives, since $G_{tt}$ and $G_{ti}$ are the Hamiltonian and momentum
constraints.

While the supersymmetrization of ${\cal L}_{I}$  has been worked out in
\cite{fer1,fer2} and ${\cal L}_{III}$ for the
 $\N=1$ case in an arbitrary Jordan frame in \cite{fer-lind}, to our
knowledge, 
the supersymmetric theory containing ${\cal L}_{II}$ was never found. It is the purpose of this
work to construct the supersymmetric version of ${\cal L}_{II}$. 

Apart from the obvious interest of studying the 
most generic supersymmetric theories avoiding Ostrogradski (higher derivatives) 
instabilities \cite{ostro,ostro2}, we note that the interaction (\ref{Gf}) 
effectively describe part of 
the cubic graviton-dilaton-dilaton vertex in heterotic
superstring theory and therefore appear in the low-energy 10D heterotic string effective
action  \cite{gross-sloan}.\footnote{However, it should also be noted that
this term has not been found in
the heterotic quartic effective supergravity action constructed in \cite{roo}.}  
Moreover, it has also been shown in \cite{maeda}, that there exists a field
redefinition up to $\alpha'$ corrections, such as to generate the terms
${\cal{L}}_I,{\cal{L}}_{II}$ out of a stringy effective action.

From a more phenomenological point of view, the theory ${\cal L}_{II}$ plays a 
fundamental role in the so called ``Gravitationally Enhanced Friction" (GEF) mechanism 
developed in \cite{germ-keh,germ-keh2,Germani:2010hd,others,others2}.
There, thanks to the
GEF, any steep (or not) scalar potential, 
can in principle produce a cosmic inflation for (relatively) small mass scale $M_{II}$. This is due to an enhanced 
friction produced by the Universe expansion acting on the (slow) rolling
scalar field. Obviously then, the 
supersymmetrization of the GEF may notably enlarge the possibilities to find inflationary scenarios in supergravity and/or 
string theory.
An additional motivation for studying supergravities with 
higher derivative terms, is related to the well known fact that they appear
in the 
 effective field theory action for the massless states of the superstring theory, after integrating out 
all superstring massive states.

All efforts to build higher-derivative supergravities in 4D are based on
off-shell formulations. 
The latter are drastically 
different from the on-shell ones and, most importantly, they are not unique. This also happens in global supersymmetry
where there are more than one off-shell formulations of an on-shell theory. We may recall for example the $\N=1$ 4D theory 
where   
a scalar and a pseudoscalar may be completed off-shell by an auxiliary scalar field
resulting in a chiral multiplet. Replacing the pseudoscalar by an 
 antisymmetric two-form,  a linear multiplet arises. In this case, there is no need of extra auxiliary fields 
as the off-shell degrees of freedom 
of an antisymmetric form field are more than those of a scalar.
These degree of freedom are the exact number needed to complete the off-shell content of the linear multiplet. On-shell, of course, the two multiplets are the same. 

This 
feature persists also in local supersymmetry where at least for the 
minimal $\N=1$ 4D supergravity we are interested in, many off-shell formulations exist. 
The reason  is that $\N=1$ superfields carry
highly reducible supersymmetry multiplets and additional constraints should be implemented 
for their truncation. Then the constraints together with
the torsion and Bianchi identities are used to solve for the independent fields. As there 
 are various ways implementing this procedure, there are also various off-shell formulations. 
Known examples are the  
off-shell supergravity  formulation 
based on the $12+12$  multiplet \cite{12old,12old2} and the new minimal
$12+12$ multiplet
\cite{12new,Sohnius:1981tp,Sohnius:1982fw,gates,Brandt:1996au}. There
are  also
other non-minimal formulations  like the one based on  the non-minimally
$20+20$ \cite{20+20,20+20b,20+20d} or $16+16$ \cite{wess,Siegel} 
multiplets. Nevertheless, these formulations may be considered reducible in the sense that 
they can be mapped to the minimal $\N=1$ 
supergravities coupled with extra multiplets. 
What is important to know though, is that it has been proven \cite{ferrara-kugo} that when no higher derivative terms are 
present,  the off-shell formulations of minimal supergravities are equivalent. 
For example old-minimal and new minimal supergravities at the 
two-derivative level are  connected by a duality transformation, 
where the chiral compensator of the former is mapped to a linear 
compensator of the latter. 
When higher derivatives are present, the duality transformation does not
work any more  due to derivatives of the compensator 
 and the two formulations are {\it not equivalent}
\cite{ferrara-kugo,Cecotti:1987qe}.

In this work we will construct the supersymmetrization of ${\cal L}_{II}$
in the
new-minimal supergravity framework of
\cite{12new,Sohnius:1981tp,Sohnius:1982fw,gates}. Our attempts in the old
minimal supergravity setup have so far failed to  
reproduce ${\cal L}_{II}$. In particular,
consideration of corresponding higher-derivative supergravity terms, like
the ones we employ here, in old minimal formulation does not seem to give
rise to such a term \cite{bauman}. Whether or not one might
nevertheless  find a way of obtaining ${\cal L}_{II}$ in the old minimal
supergravity  
formalism is an interesting open question that will not be discussed  
here but postponed for future research.

\section{New Minimal $\N=1$ 4D Supergravity}

The simplest example of $\N=1$ four-dimensional Poincar\'e supergravity is based on 12 bosonic 
and 12 fermionic off-shell degrees of freedom. These can be arranged into a multiplet in two ways. 
In the first one, the gravitational multiplet consists of 
\be
e^a_\mu\, ,~~\psi_\mu\,  , ~~b_\mu\, , ~~ M
\ee 
and describes the dynamics of the so-called old minimal (standard) supergravity. Here, $e^a_\mu$ is the
vierbein,  $\psi_\mu$ is the gravitino, $b_\mu$ is a vector, and $M$ a
scalar.
As usual the vierbein should be used to convert tangent space indices $(a,b,...)$ to world space indices $(\mu,\nu,..)$ and throughout this work the tangent space metric is mostly plus (more on conventions can be found in the appendix).

In the new minimal supergravity instead, the multiplet consists of
the vierbein $e_{\ \mu}^{a}$ and its supersymmetric
partner, the gravitino $\psi_\mu^{\alpha}$.  
In order to implement supersymmetry off-shell and the propagation of the
physical degrees of freedom only, one has to also add auxiliary fields, as
in the old minimal supergravity. However, in this case, the auxiliary
fields are no longer a vector and a scalar but 
 a 2-form $B_{\mu\nu}$ with gauge invariance (B-gauge)
 \begin{eqnarray}
\delta B_{\mu\nu}=\partial_{\mu}\xi_{\nu}-\partial_{\nu}\xi_{\mu}, \label{b-gauge}
\end{eqnarray}
and a gauge vector $A_{\mu}$ with associated R gauge invariance
\begin{eqnarray}
\delta A_{\mu}=-\partial_{\mu}\phi\ . \label{a1}
\end{eqnarray}
Thus, to wrap it up, the off-shell new minimal supergravity is based on the gravitational multiplet
\be
e^a_\mu\, ,~~\psi_\mu\,  , ~~A_\mu\, , ~~ B_{\mu\nu}\ .
\ee
For more specific details on the structure of this theory the reader should consult  \cite{ferrara-sab}. 

It has been argued that the natural superspace geometry  for four-dimensional $\N=1$ heterotic superstring 
corresponds to the new minimal formulation of the $\N=1$ supergravity
\cite{fer3,fer4,fer5}. This can be understood in terms of supervertices
and 
superspace Bianchi identities of the vertex multiplets. In addition, it seems that there is a deep connection between the $U(1)$ 
gauge symmetry of  $A_\mu$ in (\ref{a1}) above with the $U(1)$ Kac-Moody symmetry of the $\N=2$ superconformal algebra 
of the underlying superconformal theory.  

This R symmetry is however anomalous (actually it is a mixed superconformal-Weyl-$U(1)$
anomaly \cite{gates-1}). Nevertheless, by using the Green-Schwarz mechanism, the 
symmetry is restored at one loop thanks to the introduction of a matter linear multiplet together with supersymmetric Lorentz and Chern-Simons terms \cite{laurent,cardoso}.

In the new minimal supergravity, there exist three sets of chiral and Lorentz
connections
\begin{eqnarray}
\omega_{abc}^{\pm}&=&\omega_{abc}\pm H_{abc}\ ,\nonumber\\
A_\mu^{\ +}&=&A_\mu-H_\mu\ ,\\
A_\mu^{\ -}&=&A_\mu-3H_\mu\nonumber,
\end{eqnarray}
where the following notation has been used
\begin{eqnarray}
H_{\mu \nu \lambda}&=&\partial_{\mu}B_{\nu
\lambda}+\partial_{\nu}B_{\lambda \mu}+\partial_{\lambda}B_{\mu
\nu}\nonumber\\
&\ &+
\frac{i}{8}
\bar{\psi}_{\mu}\gamma_\nu\psi_\lambda+\frac{i}{8}\bar{\psi}_{\nu}
\gamma_\lambda\psi_\mu+\frac{i}{8}\bar{\psi}_{\lambda}\gamma_\mu \psi_\nu\ ,\nonumber\\
H^{\mu}&=&-\frac{1}{3!} \varepsilon^{\mu \nu \kappa \lambda} H_{\nu \kappa
\lambda}.
\end{eqnarray}
The covariant derivatives in this formulation are therefore defined as
\begin{eqnarray}
\cD&=&d+\delta_L({\cal{\omega}}_{ab})+\delta_A({\cal{A}})\ ,\nonumber\\
\cD^{\pm}&=&d+\delta_L(\omega_{ab}^{\pm})+\delta_A({\cal{A}}^{\pm}),
\end{eqnarray}
with
\begin{eqnarray}
&&\delta_A(\phi)\Phi=i\,n\, \phi\,\Phi\ , \nonumber \\
&&\delta_L(\Lambda)\Phi=\frac{i}{2}S_{ab}\Lambda^{ab}\Phi\ , \nonumber \\
&&\omega^\pm_{ab}=\omega^\pm_{ab\mu}dx^\mu\, , ~~~~{\cal{A}}^\pm=A^\pm_\mu dx^\mu\ .
\end{eqnarray}
For the gravitino, for example, we have $ S_{ab}=\sigma_{ab} /2 $ and $
n=-\gamma_5 /2 $. Here $\delta_A(\phi)$, $\delta_L(\Lambda)$
denote the $U(1)$ R-symmetry and Lorentz transformations with parameters $\phi$ and $\Lambda$, respectively. 
Supercovariant derivatives $\hat{\cD}$ are defined as usual and it should be noted for
future reference that $\hat{\cD}_a^{\ \pm} H_b=\hat{\cD}_a H_b $ and 
for any neutral vector  $\hat{\cD}_a^{\ \pm}
V^a=\hat{\cD}_a V^a $. 

The transformations of the supergravity multiplet
fields under supersymmetry are 
\cite{Sohnius:1981tp,Sohnius:1982fw,ferrara-sab}
\begin{eqnarray}
\label{newsugra-gravtrans}
\delta e_{\ \mu}^{a}&=&\frac{i}{2}\bar{\epsilon}\gamma^a\psi_\mu\ ,\nonumber\\
\delta \psi_\mu&=&-\cD_{\mu}^{\ +}\epsilon\ ,\nonumber\\
\delta
B_{\mu\nu}&=&\frac{i}{4}\bar{\epsilon}\left(\gamma_{\mu}\psi_{\nu}-\gamma_{
\nu}\psi_{\mu}\right)\ ,\\
\delta A_\mu ^{\
-}&=&\frac{i}{4}\bar{\epsilon}\gamma_{\mu}\gamma_{5}\sigma^{ab}\psi_{ab}\ ,
\nonumber
\end{eqnarray}
these transformations form an algebra along with general coordinate, Lorentz, chiral and B-gauge transformations. 
The supersymmetry parameter $\epsilon$ transforms as $\delta_A \epsilon=-(i\gamma_5/2) \phi \epsilon$ under chiral 
transformations so that 
in two component notation $\psi_\mu, \epsilon, \theta$ have chiral weight 
$\frac{1}{2}$ and $\bar{\psi}_\mu, \bar{\epsilon}, \bar{\theta}$ have chiral weight $-\frac{1}{2}$. 
The chiral weight of the other components follows by these rules. 
The gravitino curvature  used in  (\ref{newsugra-gravtrans}) is defined in
the Appendix A.
The superspace derivatives are defined in the usual way
\cite{Gates:1983nr,Grisaru1982,ferrara-sab} and the very
structure of the new minimal supergravity is incarnated in their
commutation and anti-commutation relations
\begin{eqnarray}
\label{newsugra-geometry}
\{ \nabla , \bar{\nabla} \}&=& 2i \nabla\!\!\!\!/^{-}\ ,  \nonumber \\
\left[ \nabla_a^{-} , \nabla \right] &=& \gamma_a \left( \frac{1}{2}
T^{bc} S_{bc} + T n -i  \gamma_5 E\!\!\!\!/ \nabla\right)\ , \\
\left[\nabla_a^{-} , \nabla_b^{-}\right]&=& \frac{i}{2} S^{cd} R_{cdab}^{-}
+inF_{ab}^--2E_{abc} \eta^{cd} \nabla_d^- +\frac{1}{2}\bar{T}_{ab}\nabla\ .
\nonumber
\end{eqnarray}
Here $F_{\mu\nu}=\partial_\mu A_\nu-\partial_\nu A_\mu$ is the field
strength of the gauge field $A_\mu$,
$E_{abc}=-\varepsilon_{abcd} E^d$ and the superfields $E_a$, 
$T_{ab}$ and $T$ will be 
defined in a moment.

\section{Multiplets}

A general multiplet of new minimal supergravity is 
\begin{eqnarray}
V=(C,\chi,H,K,V_a,\lambda,D).
\end{eqnarray}
It is specified by the spin and the chiral weight
\be
&&\delta_{L}C=\frac{i}{2}\Lambda_{ab}S^{ab}C\, , \\
&&\delta_{A}C=in\phi C
\ee
of its lowest component $C$, respectively. Frequently the two real scalars
$H,K$ are traded for a complex $H+iK$ one.
The supersymmetry transformations of this
multiplet are 
\begin{eqnarray}
\label{newsugra-multitrans}
\delta C &=&-\frac{1}{2}\bar{\epsilon}\chi\ , \nonumber \\
\delta \chi &=& \frac{1}{2}\left \{ i \hat{\cD\!\!\!\!/}^{-} C + \gamma_{5}
V\!\!\!\!/ +H-\gamma_{5}K\right \} \epsilon (-)^{\F}\ , \nonumber \\
\delta (H\pm iK) &=& -i \bar{\epsilon} \,\frac{1\pm \gamma_{5}}{2} \left\{
\gamma_{5} \lambda + \hat{\cD\!\!\!\!/}^{-} \chi -2i\gamma_{5} H\!\!\!\!/
\chi -i \xi C \right\}\ , \nonumber \\
\delta V_{a} &=& -\frac{i}{2} \bar{\epsilon} \left\{ \gamma_{a} \lambda -
\gamma_{5} \hat{\cD_{a}}^{-} \chi -i \gamma_{a} H\!\!\!\!/ \chi \right \}\ ,
\\
\delta \lambda &=& - \left\{ \frac{i}{4} \sigma_{ab} \hat{P}^{ab}
+\frac{i\gamma_{5}}{2} D \right\} \epsilon (-)^\F -\frac{i}{2} \xi
(\bar{\epsilon} \gamma_5 \chi)\ ,  \nonumber \\
\delta D &=& -\frac{1}{2} \bar{\epsilon} \gamma_5 \left\{
\hat{\cD\!\!\!\!/}^{-} \lambda - \gamma_a \xi V^a +i \Delta \chi \right\}\ .
\nonumber
\end{eqnarray}
We have used the following definitions
\begin{eqnarray}
\label{newsugra-page330}
\xi &=& \frac{i}{2} \psi_{ab} S^{ab} -i \gamma_5 \gamma \cdot rn\ ,   \nonumber \\
\Delta &=& -\frac{i}{2} \hat{F}_{ab}^{+} S^{ab} -\frac{i}{2} \hat{R}^- n\ , \\
 \hat{P}_{ab}  &=& \hat{\cD}_{a}^-V_b - \hat{\cD}_{b}^-V_a
-2H_{abc}V^d+\frac{i}{2} \bar{\psi}_{ab} \gamma_5 \chi\ ,  \nonumber
\end{eqnarray}
and the factor $(-)^\F$ accounts for the Fermi or Bose statistics of the
first component. Note that $\xi$ and $\Delta$ only involve the spin and
chiral generators of the first component. The properties of the general
multiplet can be encoded in the following superfield representation
\begin{eqnarray}
\label{newsugra-superfield}
&&V=C-\bar{\theta}\chi-\frac{1}{2}\bar{\theta}\left\{
H-i\gamma_5K+\gamma_5V\!\!\!\!/\right\}\theta
\ ,\nonumber\\
&&\ \ \ \ 
\  \ \ +i(\bar{\theta}\theta)\bar{\theta}\left\{\gamma_5\lambda+\frac{1 }
{ 2 } \hat { \cD\!\!\!\!/}^{-}\chi-\frac{3i\gamma_5}{2}H\!\!\!\!/\chi-i\xi
C\right\}
+\frac{1}{4}\left(\bar{\theta}\theta\right)^2\left(D+\frac{1}{2}\hat{\Box}
^-C\right).
\end{eqnarray}

Constrained multiplets may be obtained by imposing appropriate constraints
on the 
general multiplet $V$. Known representations include complex vector and
real vector multiplets, gauge and chiral multiplets
and, linear and real linear multiplets. We will discuss below the chiral
and real linear multiplets as they are 
involved in our discussion.  
 
\subsection{Chiral Multiplet}

A {\it chiral
multiplet} $\Phi(A,\chi,F)$ is defined by the constraint
$\bar{\nabla}_{\dot{\alpha}} \Phi=0 $ and its embedding in the general
multiplet is given by
\begin{eqnarray}
V(\Phi)=(A,\chi_L,F,-iF,-i\hat{\cD}_a^{-}A,-i\xi A,-i \Delta A).
\end{eqnarray}
The transformation rules are
\begin{eqnarray}
\label{newsugra-chiraltrans}
\delta A &=& \frac{1}{2} \epsilon \chi\ , \nonumber \\
(-)^{\F} \delta \chi &=& i\hat{\cD\!\!\!\!/}^{-} A \bar{\epsilon} +
F\epsilon\ , \\
\delta F &=& \frac{1}{2} \bar{\epsilon} (i\hat{\cD\!\!\!\!/}^{-} + 2H\!\!\!\!/) \chi + \bar{\epsilon} \bar{\xi} A\ , \nonumber
\end{eqnarray}
and its chiral superfield representation is
\begin{eqnarray}
\Phi=A+\theta\chi+\theta^2F.
\end{eqnarray}
Up to field redefinitions one can always define the components of a
superfield by projections.  A common projection which we use
throughout this work is \cite{Gates:1983nr,Grisaru1982}
\begin{eqnarray}
\label{newsugra-chiralprojdef}
\Phi| &=& A\ , \nonumber \\
\nabla_\alpha \Phi|&=&\chi_\alpha\ ,  \\
- \frac{1}{4} \nabla^2 \Phi| &=& F, \nonumber
\end{eqnarray}
where
 $\nabla^2\equiv\nabla^\alpha\nabla_\alpha$
and similarly $\bn^2\equiv\bn_{\dotal}\bn^{\dotal}$.  Moreover, from an arbitrary multiplet $V$ of weight $n$, one can form a chiral
multiplet with weight $n+1$ by the chiral projection operator 
\begin{eqnarray}
\Pi(V)=-\frac{1}{4}\bar{\nabla}^2V,
\end{eqnarray}
with components
\begin{eqnarray}
\Pi(V)| &=&  \bar{F}\ , \nonumber \\
\nabla_\alpha \Pi(V)|&=& i(\hat{\cD\!\!\!\!/}^{-}\bar{\chi} +2iH\!\!\!\!/
\bar{\chi} -\lambda -i \xi C)_\alpha\ , \\
- \frac{1}{4} \nabla^2 \Pi(V)|&=& \frac{1}{2}\left \{ D - i (\hat{D}^{-} -2iH ) \cdot
(V+i\hat{D}^{-}C) +i\Delta C + \frac{i}{2}
\bar{\psi}_{ab}\bar{\sigma}^{ab}\bar{\chi} +2 \bar{\xi} \bar{\chi}
\right\}\ .\nonumber
\end{eqnarray}

\subsection{Real Linear Multiplet}

A {\it real linear multiplet}  is defined by the constraints 
\be
L=L^*\, , ~~~  \nabla^2L=\bar{\nabla}^2L=0
\ee
and has zero chiral weight.
 The independent components of this multiplet are $C, \chi, V_a$, and the embedding in the general multiplet is
\begin{eqnarray}
\label{newsugra-linearembed}
V=(C,\chi,0,0,V_a,\lambda,D),
\end{eqnarray}
where the highest  $\lambda$ and $D$ components depend on the lower ones
\begin{eqnarray}
\lambda&=&-\gamma_5\hat{\cD\!\!\!\!/}^{-} \chi+2iH\!\!\!\!/ \chi -\frac{i}{2} \gamma_5 \psi_{cd} S^{cd} C\ ,  \nonumber \\
D&=& -\hat{\Box}^- C+2H\cdot V +\frac{i}{2}\bar{\psi}^{ab} (S_{ab} +\frac{\sigma_{ab}}{2}) \chi . \nonumber
\end{eqnarray}
Again one can define its independent components by projection as
\begin{eqnarray}
\label{newsugra-linearprojdef}
L| &=& C\ , \nonumber \\
\nabla_\alpha L|&=&\chi_\alpha\ , \nonumber \\
\bar{\nabla}_{\dot{\alpha}} L|&=&\bar{\chi}_{\dot{\alpha}}\ ,  \\
- \frac{1}{2} [ \nabla_{\beta},\bar{\nabla}_{\dot{\beta}}] L| &=&
V_{\beta\dot{\beta}}\ . \nonumber
\end{eqnarray}
It should be noted for future reference that one may define a new multiplet by acting with a
 superspace derivative on the general multiplet.

\subsection{Curvature Multiplets}

The {\it {gravitational curvature multiplets}} of this theory are the
Einstein
multiplet, $E_a$, and the Riemann multiplet, $T_{ab}^\alpha$. The
irreducible pieces of the Riemann multiplet are the scalar curvature
multiplet, $T^\alpha$, and the Weyl multiplet, $W_{ab}^\alpha$.
 The Einstein multiplet is a real linear multiplet (with chiral weight zero),
which means that 
\be
 E_{a}=E_{a}^{*} \, , ~~~~\nabla^2E_a=\bar{\nabla}^2E_a=0\ ,
\ee
and moreover, it satisfies the Bianchi identity 
\be
\nabla_aE^a=0\ ,
\ee
 a property that only appears in the new minimal supergravity and it is of crucial importance
for our results. Indeed, one can see that the independent components
of the Einstein multiplet contain the Einstein tensor as
the highest component. Specifically
\begin{eqnarray}
\label{newsugra-E_a}
E_a=\left(H_a, i \gamma_5 r_a, \frac{1}{2}
(\hat{G}_{ab}^+-\, ^*\hat{F}_{ab}^+) \right),
\end{eqnarray}
where $\hat{G}^+_{ab}- \, ^*\hat{F}^+_{ab}=\hat{G}_{ab}- \,
^*\hat{F}_{ab}-g_{ab}H_dH^d-2H_aH_b$ with $^*\hat{F}_{ab}^+$  the
supercovariant dual of the field strength defined as $^*
F_{\mu \nu}= \frac{1}{2} \varepsilon_{\mu \nu \kappa \lambda}F^{\kappa
\lambda}$. Moreover, $r_\mu$ is the Rarita-Schwinger
operator  and
$\hat{G}_{ab}$ is the supercovariant Einstein tensor
\cite{ferrara-sab}.
The Riemann multiplet is chiral $\left(\bn_{\dotal}T^\alpha_{ab}=0\right)$
with components 
\begin{eqnarray}
T_{ab}=\psi_{ab}-\left(\frac{i}{2}\sigma^{cd}\hat{\cal{R}}^+_{cdab}
+i\hat{F}^+_{ab}\right)\theta+i\hat{\cD\!\!\!\!/}^-\bar{\psi}_{ab}
\theta^2.
\end{eqnarray}
The rest curvature multiplets are defined as
\begin{eqnarray}
\label{newsugra-curv-mult}
T&=&\frac{1}{2}\sigma_{ab}T^{ab}\ ,\nonumber \\
W_{ab}&=&\frac{1}{24}\left(3\sigma_{cd}\sigma_{ab}+\sigma_{ab}\sigma_{cd}
\right)T^ { cd }\ , \nonumber
\end{eqnarray}
that is the scalar curvature multiplet and Weyl multiplet respectively.
Finally, there also exists the gauge multiplet of the supersymmetry algebra, namely
\begin{eqnarray}
\label{gaugemult}
V_p=\left(A_\mu^-,-\gamma_5\gamma\cdot
r,-\frac{1}{2}\hat{\cal{R}}^-\right),
\end{eqnarray}
 with $\hat{\cal{R}}^-=\hat{\cal{R}}+6H_aH^a$, which we will use in the following.

\section{Supersymmetric Actions}

Chiral multiplets with chiral weight $n=1$ can be used to form invariant
actions by the $F$-density formula  
\cite{Sohnius:1982fw}
\begin{eqnarray}
\label{Fdens2}
[\Sigma]_F=e\left\{F+\frac{i}{2}\chi\sigma\cdot\bar{\psi}+\frac{i}{2}A\,
\bar{\psi}^a\bar{\sigma}_{ab}\bar{\psi}^b \right\}.
\end{eqnarray}
In superfield notation this can be written as
\begin{eqnarray}
\label{Fterm}
 [\Sigma]_F=\int d^2\theta \,
{\cal{E}}\Sigma,
\end{eqnarray}
with 
\begin{eqnarray}
\label{ESigma}
 {\cal{E}}=e\left\{1-i\theta\sigma\cdot\bar{\psi}+\frac{i}{2}
\theta^2\bar{\psi}^a\bar{\sigma}_{ab}\bar{\psi}^b\right\}.
\end{eqnarray}
The restriction $n=1$ follows as $d\theta$ has $n=-\frac{1}{2}$ ($d\theta$
has $n=\frac{1}{2}$).
 Furthermore, one can also build invariant actions from a multiplet with
chiral weight zero, using the $D$-density formula 
 \begin{eqnarray}
\label{Ddens2}
[V]_D=
e\left\{D-\frac{1}{2}\bar{\psi}
\cdot\gamma\gamma_5\,\lambda+\left(V_\mu+\frac {i} {2}
\bar{\psi}_\mu\gamma_5\chi
\right)\varepsilon^{\mu\nu\rho\lambda}\partial_\nu
B_{\rho\lambda}\right\}+{\text{surface terms}}.
\end{eqnarray}
 We mention here that the $F$ and $D$ density formulas are related by
$[V]_D=2[\Pi(V)]_F+{\text{surface terms}}$.

The action for Poincar\'e supergravity is obtained by the Fayet-Iliopoulos term of the chiral gauge multiplet 
(\ref{gaugemult})
and reads
\be
\frac{1}{\kappa^2}{\cal{L}}_{sugra}=   \frac{1}{\kappa^2}
[V_p]_D=  \frac{1}{\kappa^2}\,e\left(\frac{1}{2}
R+\bar{\psi}^a r_a
+ 2A_a H^a-3H_a H^a\right) \label{vd}\ .
\ee  
Variation of the action (\ref{vd}) with respect to $A_\mu$ and $B_{\mu\nu}$  gives
\be
H_\mu=0=\epsilon^{\mu\nu\rho\sigma}\partial_\mu A_\nu\ .
\ee
Thus the vector $H_\mu$ vanish and $A_\mu$ reduces to a pure gauge and can 
therefore be set to zero by a gauge transformation. Finally then, the on-shell action of the new-minimal 
supergravity turns out to be
\be
S_{sugra}^{on-shell}=\frac{1}{\kappa^2}\int d^4x\, e\left(\frac{1}{2}
R+\bar{\psi}^a r_a\right)\ ,
\ee
which matches the on-shell $\N=1$ old minimal supergravity \cite{ffn,ffn2}.

\subsection{Non-Minimal Derivative Couplings}

In order to construct non-minimal derivative couplings, we will introduce a chiral superfield $\Phi$ with chiral weight $n=0$.
Since the kinetic term of a general chiral superfield is given by the $F$-term
density formula (\ref{Fterm}), we will have in our case
\begin{equation}
\label{Lk}
\mathcal{L}^{(0)}_{kin}=\int d^2\theta \,
{\cal{E}}  \Phi\left[-\frac{1}{4}\bn^2 \Phi^\dagger
\right]+ \text{h.c.}\ ,
\end{equation}
where $-\frac{1}{4}\bn^2$ is the chiral projection operator for the new
minimal supergravity. In component form, and recalling that $\Phi$ has a zero
chiral weight $n=0$, the bosonic part of the Lagrangian (\ref{Lk}) is found to be
 \begin{eqnarray}
 \label{Lk2}
\mathcal{L}^{(0)}_{kin}=2e\, A\Box A^*+2eFF^*-2ie
H^c\left(A\,\partial_c A^*-A^* \partial_c A\right)\ .
\end{eqnarray}

We should couple now the chiral multiplet $\Phi$ to some curvature multiplet in order to get the
 the desired non-minimal derivative
coupling (\ref{Gf}).  As both $\Phi$ and $E_a$ have zero
chiral weight,  the term $\Phi^\dagger E^a\nabla_a^-\Phi$ is a
general superfield with zero chiral weight as well. Therefore
$\bn^2\left[\Phi^\dagger E^a\nabla_a^-\Phi\right]$ is a chiral superfield
with chiral weight $n=1$ and thus the  superspace Lagrangian
\begin{equation}
\label{Lpre}
\mathcal{L}^{(0)}_{int}=\int d^2\theta \,
{\cal{E}}\left\{-\frac{i}{4}\bn^2\left[\Phi^\dagger
E^a\nabla_a^-\Phi\right]
\right\}+ \text{h.c.}\ .
\end{equation}
is supersymmetric. 
Now, (\ref{Lpre}) can be expanded as 
\begin{eqnarray}
\label{Lpre1}
\mathcal{L}^{(0)}_{int}=
\frac{i}{16}e\nabla^2\bn^2\left[\Phi^\dagger
E^a\nabla_a^-\Phi\right]\Big|+ \text{h.c.}=A+B+C,
\end{eqnarray}
where 
\begin{eqnarray}
&&A=\frac{i}{16}e\left[\left(\nabla^2\bn^2\Phi^\dagger\right)
E^a\nabla_a^-\Phi\right]\Big|+ \text{h.c.},\nonumber\\
&&B=\frac{i}{16}e\left[\left(\bn^2\Phi^\dagger\right)
E^a\left(\nabla^2\nabla_a^-\Phi\right)\right]\Big|+ \text{h.c.},\nonumber\\
&&C=\frac{i}{16}e\left[4\left(\nabla_\gamma\bn_{\dotga}\Phi^\dagger\right)
\left(\nabla^\gamma\bn^{\dotga}E^a\right)\left(\nabla_a^-\Phi\right)\right
]\Big|+ \text{h.c.}\ .
\end{eqnarray}
Keeping only  bosonic fields, 
  after a  straightforward
calculation we find
\begin{eqnarray}
&&A=2e
H^b 
\cD_b 
A^*\, 
H^a\cD_aA
+i e\Box A^*\, 
H^a\cD_aA+ \text{h.c.}\nonumber\\
&&B=-\frac{i}{4}e
\, F^*\,H^a\left(8iFH_a -4
\cD_a^-F\right)+ \text{h.c.}\nonumber\\
&&C=\frac{1}{2}e\,\partial^dA^*\,\partial^cA
\left(G_{dc}-\eta_{dc}
H^aH_a-2H_dH_c\right)
+i e\,\partial_bA^*\,\partial_cA\,\cD^bH^c + \text{h.c.}\label{ABC}\ .
\end{eqnarray}
In the above formulas we used that $\cD_a^- F=\partial_aF-iA^-_aF$ with
$A^-_a=A_a-3H_a$, since  $F$ has a chiral weight $n_F=-1$. Additionally,
in the above derivation one should use the helpful splitting
$\nabla^\gamma\bn^{\dotga}E^a=\frac{1}{2}\left[\nabla^\gamma,\bn^{\dotga}
\right] E^a+\frac{1}{2}\left\{\nabla^\gamma,\bn^ {\dotga}\right\}E^a$.

We see that the desired nonminimal
derivative coupling with the Einstein tensor indeed appears in $C$.  
Thus, the bosonic part of the interaction reads
 \begin{eqnarray}
 \label{Lnmdc}
\mathcal{L}^{(0)}_{int}&=&
 e \,G^{ab}
\partial_aA\, \partial_b A^*+2eFF^*
H^a A_a-2eFF^* H^aH_a+ie  H^a \left(F^* \partial_a F-F\partial_a
F^*\right)\nonumber\\
&-&e \partial_bA\, \partial^b A^*     
H_aH^a+2e H^a\partial_aA\, H^b\partial_b
A^*
-ie H_c \left(\partial_bA^*\,\cD^c\partial^b A -\partial_b
A\,\cD^c\partial^b A^*\right)\ .
\end{eqnarray}

In summary, assembling the  Lagrangians (\ref{vd},\ref{Lk},\ref{Lpre}) we
find that the bosonic sector of the
theory is 
 \begin{eqnarray}
 \label{Lf2}
\mathcal{L}_{0}&=&
\frac{1}{\kappa^2}\mathcal{L}_{sugra}+\frac{1}{2}\mathcal{L}^{(0)}_{
kin
} +w^2
\mathcal{L}^{(0)}_{int}\nonumber\\
&=&\frac{1}{\kappa^2}
\left[ \frac{1}{2}e{{\cal{R}}}+2e
H^aA_a-3eH^aH_a\right]\nonumber\\
&\,&+ eA\Box
A^*+     eFF^* -ie
H^c\left(A\,\partial_c A^*-A^* \partial_c A\right)\nonumber\\
&\,&
+w^2\left[
e\, G^{ab}\partial_b
A^*\,\partial_a A+2eFF^*
H^a A_a-2eFF^* H^aH_a\right.
\nonumber\\
&\ &\ \ \ \ \ \ \ \ +ie  H^a \left(F^* \partial_a F-F\partial_a
F^*\right) -e
\partial_bA\,\partial^b A^*     
H_aH^a\nonumber\\
&\ &\left.\ \ \ \ \ \ \ \ +2e\,H^a\partial_aA\,H^b\partial_b
A^*-ie H_c \left(\partial_bA^*\,\cD^c\partial^b A -\partial_b
A\,\cD^c\partial^b A^*\right)\right],
\end{eqnarray}
where we have introduced the dimensionful parameter $w^2=\pm M_{II}^{-2}$ and $\kappa^2=M_P^{-2}$. 

We may now integrate out the auxiliary fields to find the on-shell action.
For $w^2>0$ we may  define 
\begin{eqnarray}
\label{V0}
V^a&=& A^a\Big{(}1+\kappa^2w^2FF^*\Big{)}+   \frac{\kappa^2}{2}\Big{(}iA^*
\partial^aA\nonumber\\
&&-iA
\partial^aA^*- iw^2F
\partial^aF^*+iw^2F^*
\partial^aF\nonumber\\
&&
-iw^2\partial_bA^*\cD^a\partial^b
A+iw^2\partial_bA\,\cD^a\partial^b
A^*\Big{)},
\end{eqnarray}
in terms of which (\ref{Lf2}) is written as
 \begin{eqnarray}
 \label{Lf3}
e^{-1}\mathcal{L}_0
&=&\frac{1}{\kappa^2}
\left[ \frac{1}{2}{{\cal{R}}}+2
V^a\,H_a-3H^aH_a\right] + A\Box
A^*+     FF^*   \nonumber\\
&\,& 
+w^2\left[
\, G^{ab}\partial_b
A^*\,\partial_a A-2FF^* H^aH_a \right.\nonumber\\
&\ &\left. \ \ \ \ \ \ \ \ -
\partial_bA\, \partial^b A^*\,     
H_aH^a +2H^a\partial_aA\,H^b\partial_b
A^*\right].
\end{eqnarray}
It is important to notice here that since $A,F$ have chiral weights $n=0,-1$, respectively, $V_\mu$ transforms 
under the $U(1)$ symmetry as it should, i.e., 
\be
\delta V_\mu=\partial_\mu \phi 
\ee
and thus it is physically equivalent to $A_\mu$.  

To find the on-shell action, we should eliminate the auxiliary fields $V_\mu,B_{\mu\nu}, F$. This can be done exactly 
in the same way as in the pure supergravity case (\ref{vd}) where we find $V_\mu=H_\mu=0$. 
Similarly, the elimination of the auxiliary  $F$ of the chiral superfield is
straightforward and the bosonic part of the supersymmetric Lagrangian (\ref{Lpre}) turns out to be 
 \begin{eqnarray}
 \label{Lf4}
e^{-1}\mathcal{L}_0
&=&\frac{1}{2\kappa^2} {{\cal{R}}}  + A\Box
A^* 
+w^2 
\, G^{ab}\,\partial_a
A^*\, \partial_b A\ .
\end{eqnarray}

There is a difference when $w^2<0$. Variation with respect to $A_a$ gives
the following equation
\be
\left(\frac{1}{\kappa^2}+w^2FF^*\right)H_a=0\ .
\ee
For $w^2>0$ the only solution is $H_a=0$ and we may define $V^a$ in
(\ref{V0}) as described above. However, 
for $w^2<0$, there are two solutions: i) a supersymmetric solution $H_a=0$ and ii) a non-supersymmetric one 
 $FF^*=\frac{1}{\kappa^2 w^2}$. For the supersymmetric solution, we arrive at the bosonic part (\ref{Lf4}) of our supersymmetric
theory. On the other hand for  the no-supersymmetric solution, $A^a$ cannot anymore  be traded for $V^a$. Moreover, it 
 generates a cosmological constant as expected,  
  introducing at the same time higher derivatives. Indeed, in this case, the last term of (\ref{Lf2}) would not 
vanish leading to harmful higher-derivative interactions.

 The properties of the theory (\ref{Lf4})  have been studied in
\cite{sar,sloth}. In particular, in \cite{sloth}
 the scalar $A$ has 
been dubbed as the 
{\it Slotheon} for the reason that, generically, for a given kinetic energy, its time derivative is smaller than the same 
calculated for a canonical scalar field. 
This again proves the usefulness of this theory for Inflation, where, in
order to get an accelerated expansion of the 
primordial Universe, the scalar field should have a very small time derivative. 
In \cite{sloth} it has also been proven that spherically symmetric Black Holes cannot have slotheonic hairs and, 
finally, it has been conjectured that this theory does not violate the no-hair theorem generically.
 
We should note that the Lagrangian (\ref{Lpre}) can easily be generalized to describe more general non-minimal 
couplings of the form $V(A,A^*) G^{\mu\nu}\partial_\mu A\partial_\nu A^*$.
Indeed, we may employ a holomorphic function $W(\Phi)$ as follows
\begin{equation}
\label{Lpregen}
\mathcal{L}^{(W)}_{int}=\int d^2\theta \,
{\cal{E}}\left\{-\frac{i}{4}\bn^2\left[\bar{W}(\Phi^\dagger)
E^a\nabla_a^-W(\Phi)\right]
\right\}+ \text{h.c.}\ .
\end{equation} 
The computation of (\ref{Lpregen}) goes straightforward as in the previous case and the result, after combining with (\ref{vd},\ref{Lk},\ref{Lpre}) and by doing an appropriate shifting of the $U(1)$ vector, turns out to be
\begin{eqnarray}
 \label{Lfw}
e^{-1}\mathcal{L}^{(W)}
&=&\frac{1}{\kappa^2}
\left[ \frac{1}{2}{{\cal{R}}}+2
V^a\,H_a-3H^aH_a\right] + A\Box
A^*+     FF^* \nonumber\\
&\,&
+w^2\Big{|}\frac{\partial {\cal W}}{\partial A}\Big{|}^2\left(\phantom{\frac{\cal W}{A}}\!\!\!\!\!\!\!\!\!\!
\, G^{ab}\partial_b
A^*\partial_a A-2FF^* H^aH_a\right.\nonumber\\
&\ &\left. \ \ \ \ \ \ \ \ \ \ \ \ \ \ \ \ \ \  - 
\partial_bA \partial^b A^*     
H_aH^a +2H^aH^b\partial_aA\partial_b
A^*\right)\ ,
\end{eqnarray}
where $\cal W$ is the lowest component of $W$.

Again, field equations for $V^\mu$ and  $H^\mu$ force the latter to vanish and the former to be a pure gauge. With this in mind,  
the bosonic part of the Lagrangian, after elimination of the auxiliary fields is
\begin{eqnarray}
 \label{Lf41}
e^{-1}\mathcal{L}^{(W)}
&=&\frac{1}{2\kappa^2} {{\cal{R}}}  + A\Box
A^* 
+w^2 \Big{|}\frac{\partial{ \cal W}}{\partial A}\Big{|}^2
\, G^{\mu\nu}\,\partial_\mu
A^*\, \partial_\nu A\ .
\end{eqnarray}

An obvious question concerns  possible potential terms. 
Due to the requirement of   R-invariance, one cannot use the F-density
formula (\ref{Fterm}) to write general 
Lagrangians, unless the
F-density has a total chiral weight of $n=1$. For the neutral chiral multiplet we 
have used to construct our theory, it is not possible to write an R-symmetric potential term, 
unless new chiral fields are introduced. However, one can introduce 
explicit  soft supersymmetry breaking terms  of the form
$m^2 AA^*$, as potential for the neutral scalar.

A second question is why the neutral $n=0$ prescription in (\ref{Lpre}) is fundamental to avoid higher-derivatives.
An R-charged multiplet
with $n\neq 0$ would give charge to the scalar $A$. In this case, $A$ would be minimally
coupled to the $U(1)$ gauge field
$A_\mu$ inducing quadratic terms for the gauge field. Moreover, 
 kinetic terms for both $A_\mu$ and $H_\mu$ will appear. In this case,
$A_\mu$ and $H_\mu$ could not be eliminated algebraically anymore.
Specifically, the equation for $A_\mu$ would read $H_\mu\sim \partial_\mu
A+\ldots$. 
It is then clear that the elimination of $H_\mu$ would produce
quartic derivatives of the scalar $A$ and consequently  a higher derivative theory from, 
for example, the last term of (\ref{Lf2}). Therefore, only for a neutral $n=0$ chiral field
a  theory with no harmful  higher derivatives can be obtained.
However, for completeness, in Appendix \ref{Lagrn} we present the bosonic
sector of a general R-charged chiral multiplet of chiral weight
$n$.

Finally, we note that in the
fermionic sector of the theory, among the various fermionic
interactions that  arise,
the term
\begin{equation}
 \label{Lfermionic}
 {\cal{L}}_\chi=-w^2\,e\frac{i}{4}\hat{G}^{ab}
\chi \sigma_ {b}\hD^-_a\bar{\chi}-i w^2 e  {\cal
D}_d A^* {\cal D}_a\chi \sigma^d \bar r^a\ ,
\end{equation}
is  the direct supersymmetric partner of the Einstein coupling in (\ref{Lf4}) needed to cancel scalar supersymmetry
variations of 
${\cal{L}}_{II}$ . The 
first term in (\ref{Lfermionic}) was for first time introduced in non supersymmetric models in 
\cite{Germani:2011cv}. In \cite{Germani:2011cv} it has been shown that each time couplings of the form 
(\ref{Lfermionic}) or (\ref{Lf4}) are introduced, dependently upon the scale $w$, fields get dynamically localized around domain walls.

\section{Conclusions}

General Relativity (GR) minimally coupled to scalars is not the most generic tensor-scalar theory propagating only 
a massless spin-2 and a spin-0 field. In fact, non-minimally coupled theories of curvatures to matter fields can 
also be constructed with these properties. Such theories,
which do not produce any higher derivatives in the equations of 
motion and, at the same time, maintain the GR constrains able to reduce the
graviton degrees of freedom to 
only two polarizations (in four dimensions), are found in \cite{hord}.
These theories have non-derivative and derivative 
couplings of matter to curvatures.

The supergravity extension of the non-derivatively coupled theories  such
as ${\cal{L}}_{III}$ 
has been already constructed in the literature \cite{fer-lind}. 
However, non-minimal derivative coupled supergravities to matter fields, without extra propagating 
modes, are restricted to the Gauss-Bonnet interactions
${\cal{L}}_I$.   
Here we focused on the supersymmetrization of the non-minimal derivative coupled Lagrangian,
${\cal{L}}_{II}$. This was achieved in the framework of new-minimal supergravity by employing a chiral multiplet and the 
linear curvature  multiplet. 

A theory described by (\ref{Lf4}) or, more generically, (\ref{Lf41}), may have many 
phenomenological interesting properties. 
The first one is that, each time a domain wall is present in the theory, dependently upon the scale $w$, the scalar field gets dynamically 
localized around the domain wall itself \cite{Germani:2011cv}. In fact, one may consider ${\cal{L}}_{II}$ as 
a field theoretical realization of the quasi-localization mechanism of \cite{gia}.
A second, perhaps more important, phenomenological aspect is related to Inflation.
 Whenever the background Einstein tensor is larger than the mass scale $w^{-2}$, no matter what potential is driving $A$, 
Inflation is naturally produced without exceeding the perturbative cut-off
scale of the theory, which is below the Planck scale as it should be for a
ghost-free  
theory \cite{dvali-germani}.
This is due to an enhanced gravitational friction acting on the evolving scalar field and sourced by the 
Universe expansion itself
\cite{germ-keh,germ-keh2,Germani:2010hd,others,others2}. 
We therefore believe that the supersymmetrization of the 
${\cal{L}}_{II}$ might open new possibilities for exploring inflation in
supergravity/string theory.

 \begin{acknowledgments}
The authors would like to thank G. Dvali, D. Green, S. Katmadas,  L. Martucci, J. Russo,  E. Sezgin and Y. Watanabe  for
useful discussion and correspondence. FF wishes to thank LMU University for the hospitality
during part of the preparation of the present work. CG is supported by Alexander Von Humboldt Foundation.

\end{acknowledgments}

 \appendix

\section{Conventions}

\label{conv}

Throughout the work we use a Minkowski metric with signature (-,+,+,+),
and the fully antisymmetric tensor is taken as  $\varepsilon_{0123}=+1$.
The Dirac matrix conventions are $\{\gamma_a,\gamma_b\}=-2g_{ab}$,
$\gamma_5=-i\gamma^0\gamma^1\gamma^2\gamma^3$, while we use
$\sigma_{ab}=\frac{i}{2}[\gamma_a,\gamma_b]$,
  and $\bar{\psi}=\psi^\dagger C$.  

In a Majorana representation $C =\gamma^0$ and the Majorana condition is
$\psi=\psi^\star$. The two-component spinor formalism is derived
from the following chiral
representation of the Dirac matrices,
\begin{eqnarray}
&&\gamma_{5}=\left(\begin{tabular}{lr}
    $-1$ & $\ 0$  \\
    $\ \  0$ & $\ 1$  \\
\end{tabular}\right),\ \
\gamma_{a}=\left(\begin{tabular}{rr}
    $0\ \, $ & $\sigma_a$  \\
    $\bar{\sigma}_a$ & 0  \\
\end{tabular}\right)
\ ,\nonumber\\
&&\sigma_a=(1,\vec{\sigma}),\ \ \ \bar{\sigma}_a=(1,-\vec{\sigma})\ ,
 \nonumber\\
&&\psi=\left(\begin{tabular}{r}
   $\psi_\alpha$   \\
    $\bar{\psi}^{\dotal}$  \\
\end{tabular}\right),\ \ 
\bar{\psi}=\left( -\psi^\alpha, -\bar{\psi}_{\dotal}\right)\ .
\end{eqnarray}
The gravitino curvature is given by
\be
\psi_{\mu\nu}=\cD^+_\mu\psi_\nu-\cD^+_\nu\psi_\mu\, , ~~~~~\psi_{ab}=e_\mu^a e_\nu^b\psi_{\mu\nu}
\ee
 and the Rarita-Schwinger operator is
\be
r^a=\frac{1}{4}\gamma_5\gamma_b\varepsilon^{bade}\psi_{de}\ .
\ee
Finally,
the Ricci scalar, the Ricci tensor and the Riemann curvature are given by
 \begin{eqnarray}
{\cal{R}}&=&\eta^{ca}{\cal{R}}_{ca}\ ,\nonumber\\
{\cal{R}}_{ca}&=&{\cal{R}}_{nma}^{\ \ \ \ b}\,e_b^{\ m}e_c^{\ n}\ ,\nonumber\\
{\cal{R}}_{nma}^{\ \ \ \ b}&=&\partial_n\omega_{ma}^{\ \
b}-\partial_n\omega_{na}^{\ \ b}+\omega_{ma}^{\ \ c}\omega_{nc}^{\ \
b}-\omega_{na}^{\ \ c}\omega_{mc}^{\ \ b}\ .
\end{eqnarray}

\section{Lagrangian for non-zero chiral weight}
\label{Lagrn}

 The bosonic part of the Lagrangian for a general chiral weight $n$ reads:
 \begin{eqnarray}
 \label{Lffinal}
e^{-1}\mathcal{L}_n
&=&
\frac{1}{\kappa^2}
\left[ \frac{1}{2}{{\cal{R}}}+2
H^aA_a-3 H^aH_a\right]\nonumber\\
&\,&+ A\Box^-
A^*+     FF^* -\frac{1}{2}n\,AA^\star\left({\cal{R}}+6H^aH_a\right) -i
H^c\left(A\,\cD_c^- A^*-A^* \cD_c^- A\right)\nonumber\\
&\,&
+w^2
\left\{i H^b \left[ \Box^- A^* \cD_{b}^- A - \Box^- A \cD_{b}^- A^* \right]
+
\frac{i}{2}
n
H^b \left({\cal{R}} + 6 H^aH_a\right) ( A \cD_{b}^- A^* - A^* \cD_{b}^- A
) \right.\nonumber \\
&&\ \ \ \ \ \ \ \ \,
+ 4 H^c
\cD_{c}^- A^* H^b \cD_{b}^- A   +i
\cD_{d}^-
A^* \cD_{a}^- A\left( \cD^{d} H^a - \cD^{a} H^d \right) 
\nonumber \\
&&\ \ \ \ \ \ \ \ \, + \cD_{d}^- A^* \cD_{a}^- A \left[ G^{da} - g^{da}
H^bH_b -2 H^d H^a \right]\nonumber \\
&&\ \ \ \ \ \ \ \ \,+ i H^a \left( F^* \cD_{a}^- F - F \cD_{a}^- F^*
\right) + 4 F F^* H^aH_a\nonumber \\
&&\ \ \ \ \ \ \ \ \, +\frac{i}{2} n H^d {\cal{R}} \left( A \cD_{d}^- A^* -
A^* \cD_{d}^- A \right) + 3
i n\,
H^aH_a H^d ( A \cD_{d}^- A^* - A^* \cD_{d}^- A  ) \nonumber \\
&&\ \ \ \ \ \ \ \ \,  + i n \ ^* F^{da} H_a ( A \cD_{d}^- A^* - A^*
\cD_{d}^- A  ) + i n\, H_a
(\cD_l H_b ) {\cal{\varepsilon}}^{blda}  ( A \cD_{d}^- A^* - A^* \cD_{d}^-
A 
)
\nonumber \\
&&\ \ \ \ \ \ \ \ \, - \frac{1}{2} n H_a \ ^*F_{lb} \,\varepsilon^{blda}  
\left( A \cD_{d}^-
A^*
+ A^* \cD_{d}^- A  \right) -n H^l (\cD_l H^d ) \left( A \cD_{d}^- A^* + A^*
\cD_{d}^-
A 
\right)  \nonumber \\
&&\ \ \ \ \ \ \ \ \, + n H^b (\cD^d H_b ) ( A \cD_{d}^- A^* + A^* \cD_{d}^-
A  )            
\nonumber \\
&&\ \ \ \ \ \ \ \ \, - \frac{1}{4} n AA^* \left( {\cal{R}} + 6 H^aH_a  
\right)^2  - \frac{1}{2} n
AA^* \
^*F_{dc} \ ^*F^{dc} - n AA^* \varepsilon^{cdka} \ ^*F_{ka} \cD_d H_c
\nonumber \\
&&\ \ \ \ \ \ \ \ \, \left. + n AA^*  (\cD_d H_c)\left( \cD^d H^c - \cD^c
H^d  \right)
\right\}\ .
\end{eqnarray}
It is clear that, for $n\neq 0$, the vector $A_a$ of field strength 
$F^{ab}$ becomes dynamical and therefore, as discussed in the text, cannot be removed by a gauge transformation.


\begin{thebibliography}{99}

\bibitem {hord}
G. W. Horndeski, 
  {\it{Second-order scalar-tensor field equations in a four-dimensional
space}},
Int. J. Theor. Phys. 10, 363 (1974).

  \bibitem{adm}
  C.~W.~Misner, K.~S.~Thorne and J.~A.~Wheeler,
   {\it{Gravitation}},
  San Francisco 1973.


\bibitem{fer1} 
  S.~Cecotti, S.~Ferrara, L.~Girardello and M.~Porrati,
   {\it{Lorentz Chern-simons Terms In N=1 Four-dimensional Supergravity
Consistent With Supersymmetry And String Compactification}},
  Phys.\ Lett.\ B {\bf 164}, 46 (1985).

\bibitem{fer2} 
  S.~Cecotti, S.~Ferrara, L.~Girardello, A.~Pasquinucci and M.~Porrati,
   {\it{Matter Coupled Supergravity With Gauss-bonnet Invariants: Component
Lagrangian And Supersymmetry Breaking}},
  Int.\ J.\ Mod.\ Phys.\ A {\bf 3}, 1675 (1988).


\bibitem{fer-lind} 
  S.~Ferrara, R.~Kallosh, A.~Linde, A.~Marrani and A.~Van Proeyen,
   {\it{Jordan Frame Supergravity and Inflation in NMSSM}},
  Phys.\ Rev.\ D {\bf 82}, 045003 (2010)
  [\href{http://xxx.lanl.gov/abs/1004.0712}
{{\tt arXiv:1004.0712}}].

 
\bibitem{ostro}
M.~Ostrogradski,
  {\it{ 1850 Memoires sur les equations differentielles relatives
au probleme des isoperimetres}},
 Mem. Ac. St. Petersburg, VI Series,
Vol. 4 385.

\bibitem{ostro2} 
 R.~P.~Woodard,
   {\it{Avoiding dark energy with 1/r modifications of gravity}},
  Lect.\ Notes Phys.\  {\bf 720 } (2007)  403,
[\href{http://xxx.lanl.gov/abs/astro-ph/0601672}
{{\tt arXiv:astro-ph/0601672}}].
 
\bibitem{gross-sloan} 
  D.~J.~Gross and J.~H.~Sloan,
   {\it{The Quartic Effective Action for the Heterotic String}},
  Nucl.\ Phys.\ B {\bf 291}, 41 (1987).

\bibitem{roo} 
  E.~A.~Bergshoeff and M.~de Roo,
   {\it{The Quartic Effective Action Of The Heterotic String And
Supersymmetry}},
  Nucl.\ Phys.\ B {\bf 328}, 439 (1989).
 
\bibitem{maeda} 
  K.~-i.~Maeda, N.~Ohta and R.~Wakebe,
   {\it{Accelerating Universes in String Theory via Field Redefinition}},
  [\href{http://xxx.lanl.gov/abs/1111.3251}
{{\tt arXiv:1111.3251}}].


\bibitem{germ-keh}
  C.~Germani and A.~Kehagias,
   {\it{New Model of Inflation with Non-minimal Derivative Coupling
 of Standard
  Model Higgs Boson to Gravity}},
  Phys.\ Rev.\ Lett.\  {\bf 105}, 011302 (2010),
  [\href{http://xxx.lanl.gov/abs/1003.2635}
{{\tt arXiv:1003.2635}}].
 

\bibitem{germ-keh2}
  C.~Germani, A.~Kehagias,
   {\it{Cosmological Perturbations in the New Higgs Inflation}},
  JCAP {\bf 1005}, 019 (2010),
  [\href{http://xxx.lanl.gov/abs/1003.4285}
{{\tt arXiv:1003.4285}}].

\bibitem{Germani:2010hd} 
  C.~Germani and A.~Kehagias,
   {\it{UV-Protected Inflation}},
  Phys.\ Rev.\ Lett.\  {\bf 106}, 161302 (2011)
  [\href{http://xxx.lanl.gov/abs/1012.0853}
{{\tt arXiv:1012.0853}}].

 

 \bibitem{others}
 C.~Germani and Y.~Watanabe,
 {\it{UV-protected (Natural) Inflation: Primordial Fluctuations and
non-Gaussian Features}},
  JCAP {\bf 1107} (2011) 031,
  [\href{http://xxx.lanl.gov/abs/1106.0502}
{{\tt arXiv:1106.0502}}].
   
  \bibitem{others2}
  C.~Germani,
 {\it{Slow Roll Inflation: A Somehow Different Perspective}},
  [\href{http://xxx.lanl.gov/abs/1112.1083}
{{\tt arXiv:1112.1083}}]
 

\bibitem{12new} 
  V.~P.~Akulov, D.~V.~Volkov and V.~A.~Soroka,
   {\it{On the General Covariant Theory of Calibrating Poles in
Superspace}},
  Theor.\ Math.\ Phys.\  {\bf 31}, 285 (1977).
   
\bibitem{Sohnius:1981tp} 
  M.~Sohnius and P.~C.~West,
   {\it{An Alternative Minimal Off-Shell Version of N=1 Supergravity}},
  Phys.\ Lett.\ B {\bf 105}, 353 (1981).
  
\bibitem{Sohnius:1982fw} 
  M.~Sohnius and P.~C.~West,
   {\it{The Tensor Calculus And Matter Coupling Of The Alternative Minimal
 Auxiliary Field Formulation Of N=1 Supergravity}},
  Nucl.\ Phys.\ B {\bf 198}, 493 (1982).

 
\bibitem{gates}
 S.~J.~Gates, Jr., M.~Rocek and W.~Siegel,
   {\it{Solution to constraints for n=0 supergravity}},
  Nucl.\ Phys.\ B {\bf 198}, 113 (1982).
 

 
\bibitem{12old2} 
  S.~Ferrara and P.~van Nieuwenhuizen,
   {\it{The Auxiliary Fields of Supergravity}},
  Phys.\ Lett.\ B {\bf 74}, 333 (1978).

\bibitem{12old} 
 K.~S.~Stelle and P.~C.~West,
   {\it{Minimal Auxiliary Fields for Supergravity}},
  Phys.\ Lett.\ B {\bf 74}, 330 (1978).

\bibitem{Brandt:1996au} 
  F.~Brandt,
   {\it{Local BRST cohomology in minimal D = 4, N=1 supergravity}},
  Annals Phys.\  {\bf 259}, 253 (1997),
  [\href{http://xxx.lanl.gov/abs/hep-th/9609192}
{{\tt arXiv:hep-th/9609192}}].
 
\bibitem{20+20} 
  P.~Breitenlohner,
   {\it{A Geometric Interpretation of Local Supersymmetry}},
  Phys.\ Lett.\ B {\bf 67}, 49 (1977).
 
 
\bibitem{20+20b} 
  P.~Breitenlohner,
   {\it{Some Invariant Lagrangians for Local Supersymmetry}},
  Nucl.\ Phys.\ B {\bf 124}, 500 (1977).

\bibitem{20+20d} 
  W.~Siegel and S.~J.~Gates, Jr.,
   {\it{Superfield Supergravity}},
  Nucl.\ Phys.\ B {\bf 147}, 77 (1979).

\bibitem{wess} 
  G.~Girardi, R.~Grimm, M.~Muller and J.~Wess,
   {\it{Antisymmetric Tensor Gauge Potential In Curved Superspace And A
(16+16) Supergravity Multiplet}},
  Phys.\ Lett.\ B {\bf 147}, 81 (1984).

\bibitem{Siegel} 
 W.~Siegel,
   {\it{16/16 Supergravity}},
  Class.\ Quant.\ Grav.\  {\bf 3}, L47 (1986).


\bibitem{ferrara-kugo} 
  S.~Ferrara, L.~Girardello, T.~Kugo and A.~Van Proeyen,
   {\it{Relation Between Different Auxiliary Field Formulations Of N=1
Supergravity Coupled To Matter}},
  Nucl.\ Phys.\ B {\bf 223}, 191 (1983).

\bibitem{Cecotti:1987qe} 
  S.~Cecotti, S.~Ferrara, M.~Porrati and S.~Sabharwal,
   {\it{New Minimal Higher Derivative Supergravity Coupled To Matter}},
  Nucl.\ Phys.\ B {\bf 306}, 160 (1988).

\bibitem{bauman}
D.~Baumann and D.~Green,
  {\it{Supergravity for Effective Theories}},
  JHEP {\bf 1203}, 001 (2012),
  [\href{http://xxx.lanl.gov/abs/1109.0293}
{{\tt arXiv:1109.0293}}].

\bibitem{ferrara-sab}
 S.~Ferrara and S.~Sabharwal,
   {\it{Structure Of New Minimal Supergravity}},
  Annals Phys.\  {\bf 189}, 318 (1989).

\bibitem{fer3} 
  S.~Cecotti, S.~Ferrara and L.~Girardello,
   {\it{Gravitational Supervertices In N=1, 4-d Superstrings}},
  Phys.\ Lett.\ B {\bf 198}, 336 (1987).

\bibitem{fer4} 
 J.~Lauer, D.~Lust and S.~Theisen,
   {\it{Four-dimensional Supergravity From Four-dimensional Strings}},
  Nucl.\ Phys.\ B {\bf 304}, 236 (1988).

\bibitem{fer5} 
 B.~A.~Ovrut,
   {\it{Superstrings And The Auxiliary Field Structure Of Supergravity}},
  Phys.\ Lett.\ B {\bf 205}, 455 (1988).

\bibitem{gates-1} 
  S.~J.~Gates, Jr., M.~T.~Grisaru and W.~Siegel,
   {\it{Auxiliary Field Anomalies}},
  Nucl.\ Phys.\ B {\bf 203}, 189 (1982).

\bibitem{laurent} 
  L.~Baulieu and M.~P.~Bellon,
   {\it{Anomaly Cancellation Mechanism In N=1, D = 4 Supergravity And
Distorted Supergravity With Chern-simons Forms}},
  Phys.\ Lett.\ B {\bf 169}, 59 (1986).

\bibitem{cardoso} 
  G.~Lopes Cardoso and B.~A.~Ovrut,
   {\it{A Green-Schwarz mechanism for D = 4, N=1 supergravity anomalies}},
  Nucl.\ Phys.\ B {\bf 369}, 351 (1992).


\bibitem{Gates:1983nr} 
  S.~J.~Gates, M.~T.~Grisaru, M.~Rocek and W.~Siegel,
   {\it{Superspace Or One Thousand and One Lessons in Supersymmetry}},
  Front.\ Phys.\  {\bf 58}, 1 (1983)
  [\href{http://xxx.lanl.gov/abs/hep-th/0108200}
{{\tt arXiv:hep-th/0108200}}].

\bibitem{Grisaru1982} 
M.T. Grisaru in
  {\it{Supersymmetry And Supergravity '82. Proceedings, School, Trieste,
Italy, September 6-15, 1982,}}
  Singapore, Singapore: World Scientific (1983).


\bibitem{ffn} 
  D.~Z.~Freedman, P.~van Nieuwenhuizen and S.~Ferrara,
   {\it{Progress Toward a Theory of Supergravity}},
  Phys.\ Rev.\ D {\bf 13}, 3214 (1976).

\bibitem{ffn2} 
  S.~Deser and B.~Zumino,
   {\it{Consistent Supergravity}},
  Phys.\ Lett.\ B {\bf 62}, 335 (1976).
 


\bibitem{sar}
 E.~N.~Saridakis and S.~V.~Sushkov,
   {\it{Quintessence and phantom cosmology with non-minimal derivative
coupling}},
 Phys.\ Rev.\ D {\bf 81}, 083510 (2010),
  [\href{http://xxx.lanl.gov/abs/1002.3478}
{{\tt arXiv:1002.3478}}].

 

\bibitem{sloth}
C.~Germani, L.~Martucci and P.~Moyassari,
   {\it{Introducing the Slotheon: a slow Galileon scalar field in curved
space-time}},
  [\href{http://xxx.lanl.gov/abs/1108.1406}
{{\tt arXiv:1108.1406}}].

  
\bibitem{Germani:2011cv} 
  C.~Germani,
   {\it{Spontaneous Localization on a Brane: a High-Energy Resolution of
Braneworlds}},
  [\href{http://xxx.lanl.gov/abs/1109.3718}
{{\tt arXiv:1109.3718}}].
 
 
   
\bibitem{gia}
G.~R.~Dvali, G.~Gabadadze and M.~A.~Shifman,
  {\it{(Quasi)localized gauge field on a brane: Dissipating cosmic
radiation to extra dimensions?}},
  Phys.\ Lett.\ B {\bf 497} (2001) 271,
  [\href{http://xxx.lanl.gov/abs/hep-th/0010071}
{{\tt arXiv:hep-th/0010071}}].
 

\bibitem{dvali-germani}
G.~Dvali, S.~Folkerts and C.~Germani,
   {\it{Physics of Trans-Planckian Gravity}},
   Phys.\ Rev.\ D {\bf 84} (2011) 024039,
  [\href{http://xxx.lanl.gov/abs/1006.0984}
{{\tt arXiv:1006.0984}}].

 

\end{thebibliography}
\end{document}